\newcommand{\CLASSINPUTtoptextmargin}{2cm}
\newcommand{\CLASSINPUTbottomtextmargin}{2.55cm}
\documentclass[conference]{IEEEtran}
\IEEEoverridecommandlockouts
\usepackage{cite}
\usepackage{amsmath,amssymb,amsfonts}
\usepackage{textcomp}
\usepackage{bm}
\usepackage{color, xcolor}
\usepackage{enumerate}
\usepackage{enumitem}
\usepackage{mathrsfs}
\usepackage{graphicx}
\usepackage{multirow}
\newcommand{\BS}{\begin{subequations}}
\newcommand{\ES}{\end{subequations}}
\newcommand{\BE}{\begin{equation}}
\newcommand{\EE}{\end{equation}}
\newtheorem{theorem}{Theorem}

\newtheorem{assumption}{Assumption}

\newtheorem{remark}{Remark}
\newtheorem{lemma}{Lemma}
\newtheorem{corollary}{Corollary}

\begin{document}
\title{Capacity-Region-Achieving Sparse Regression  Codes for MIMO Multiple-Access Channels
\thanks{The work of H. Yan and L. Liu was supported in part by the National Natural Science Foundation of China (NSFC) under Grants 62301485 and 62394292, in part by the Zhejiang Provincial Natural Science Foundation under Grant LZ25F010002, and in part by the State Key Laboratory of Integrated Services Networks under Grant ISN25-10. The work of G. Caire and B. \c{C}akmak was supported by the Gottfried Wilhelm Leibniz-Preis 2021 of the German Science Foundation (DFG).}
}

\author{
\IEEEauthorblockN{Hao~Yan$^{*\dagger}$,~Lei~Liu$^{*\dagger}$,~Yuhao~Liu$^\ddagger$,~Burak~\c{C}akmak$^\S$,~Giuseppe~Caire$^\S$}
  \IEEEauthorblockA{$^*$College of Information Science and Electronic Engineering, Zhejiang University, China\\
  $\dagger$State Key Laboratory of Integrated Services Networks, Xidian University, China\\
  $^\ddagger$Department of Mathematical Science, Tsinghua University, China\\
  $^\S$Faculty of Electrical Engineering and Computer Science, Technical University of Berlin, Germany \\
  Email: \{hao\_yan, lei\_liu\}@zju.edu.cn, yh-liu21@mails.tsinghua.edu.cn, \{burak.cakmak, caire\}@tu-berlin.de}\\
}
\maketitle

\begin{abstract}
      This paper proposes a coding framework for capacity-region-achieving sparse regression (SR) codes over MIMO multiple-access channels (MIMO-MAC), where a single SR code is used for each user at the transmitter. With random semi-unitary dictionary matrices applied for encoding, multiple-access OAMP (MA-OAMP) enables reliable parallel interference cancellation (PIC) at the receiver. Theoretically, an optimal coding principle with the MA-OAMP receiver, which achieves the sum capacity and, in combination with time sharing, achieves the entire capacity region, is established as the guiding principle for designing capacity-region-achieving codes. Accordingly, a coding scheme for capacity-region-achieving SR codes is proposed via proper power allocation over the position-modulated signals. 
\end{abstract}

\section{Introduction}

Channel capacity defines the fundamental limit of reliable information transmission and serves as the theoretical benchmark for coding design. Over the decades, numerous forward error correction (FEC) coding schemes, such as Turbo codes\cite{Berrou1993Turbo}, low-density parity-check (LDPC) codes\cite{Richardson2001,Kudekar2011}, and Polar codes\cite{Arikan2009}, have been developed to approach this limit for discrete memoryless channels (e.g., binary symmetric channels (BSC), binary erasure channels (BEC) and binary-input AWGN), with LDPC and Polar codes now widely adopted in 5G standards. More recently, sparse regression (SR) codes\cite{Joseph2012,Joseph2014} have emerged as a simple yet powerful coding scheme capable of achieving AWGN channel capacity when decoded via approximate message passing (AMP)\cite{Donoho2009,Donoho2010} for i.i.d. Gaussian dictionary matrices\cite{Rush2017,Rush2019,Barbier2017,Ramji2019}, and via its variants, including orthogonal AMP (OAMP)\cite{Ma2017} and vector AMP (VAMP)\cite{Rangan2019vector}, for right-unitarily invariant dictionary matrices\cite{Hou2022,Xu2023,Liu2025CapacityachievingSS}. Originally, SR codes were developed for memoryless channels, such as point-to-point AWGN channels \cite{Rush2017,Rush2019,Barbier2017,Hou2022,Xu2023,Liu2025CapacityachievingSS}, Gaussian multiple-access channels (MAC) \cite{Ramji2019,Fengler2021}, and Gaussian broadcast channels \cite{Ramji2019}. However, practical communication environments often involve more complex multiple-input multiple-output (MIMO) channels, typically including multipath channels and MIMO multiple-access channels (MIMO-MAC), where AWGN-oriented coding strategies suffer performance degradation and fail to achieve the capacity or capacity region\cite{Liu2019,Wang2020,Liu2021,Chi2022,Liu2024}. In contrast, AMP-based optimal coding schemes were proposed \cite{Liu2021,Liu2024} for the design of capacity or capacity-region-achieving codes over MIMO channels \cite{Liu2021,Liu2024,Chi2022} when decoding with AMP-type algorithms\cite{Donoho2009,Ma2017,Rangan2019vector,Liu2022memory}. However, AMP-type algorithms require the observation matrix to satisfy strict conditions (e.g., i.i.d. Gaussian entries or right-unitary invariance) that rarely hold in practical channels. While the multi-layer superposition coding modulation (SCM) in \cite{Liu2021} follows this principle, it is hard to implement owing to the complexity issue, as it requires the design and successive interference cancellation (SIC) of multiple layers with distinct AWGN-capacity-achieving codes. To the best of our knowledge, no capacity or capacity-region-achieving coding scheme exists that uses a single code for each user while decoding with parallel interference cancellation (PIC) in MIMO channels.

To tackle these issues, we present a capacity-region-achieving coding framework over MIMO-MAC using a single SR code for each user (i.e., without employing SCM) at the transmitter. By encoding with random semi-unitary dictionary matrices, the MA-OAMP receiver enables Bayes-optimal PIC\cite{Burak2026}. Accordingly, the achievable user-rate suprema can be expressed in analytical forms, which yield an optimal coding principle that achieves the entire capacity region. As a concrete realization, we design SR codes through proper power allocation over position-modulated signals. Numerical results show that the proposed SR codes can obtain near capacity-region-achieving performance in MIMO-MAC.

\section{Transceiver Model}
\subsection{MIMO-MAC with SR Codes}
Consider a MIMO-MAC with $U$ users, in which each user transmits an SR codeword $\mathbf{x}_u$. The received signal is given by
\begin{align}
\mathbf{y}
=\sum_{u=1}^U \mathbf{H}_u \mathbf{x}_u + \mathbf{n}
=\sum_{u=1}^U \mathbf{H}_u \mathbf{\Xi}_u \mathbf{s}_u + \mathbf{n},
\label{equ: MIMO-MAC with mu-RT}
\end{align}
with $\mathbf{n}\sim \mathcal{CN}(\mathbf{0},\sigma^2 \mathbf{I})$. Here, $\mathbf{x}_u \in \mathbb{C}^{M_u}$ is an SR codeword obtained by applying a random semi-unitary dictionary matrix $\mathbf{\Xi}_u \in \mathbb{C}^{M_u \times N_u}$ to a position-modulated signal $\mathbf{s}_u \in \mathbb{C}^{N_u}$, which consists of $L_u$ independent sections of dimension $B_u$ with exactly one non-zero entry $\{\sqrt{p_{u,l}}\}_{l=1}^{L_u}$ per section. The SR codewords satisfy a unit average power constraint $\sum_{l=1}^{L_u} p_{u,l}/N_u = 1$. By choosing $\mathbf{\Xi}_u$ as a deterministic realization of, for example, Haar, i.i.d., or permutation invariant matrices \cite{Dudeja2024,Lei2025Random}, the equivalent channel $\mathbf{A}_u = \mathbf{H}_u \mathbf{\Xi}_u$ is a deterministic realization of the universality class\cite{Dudeja2024}, enabling AMP-type algorithms to achieve Bayes-optimal performance. Throughout this paper, the receiver is assumed to have perfect knowledge of $\{\mathbf{H}_u\}_{u=1}^U$. We consider static $\mathbf{H}_u \in \mathbb{C}^{M \times M_u}$ with $M = \bar{M} M_{\rm R}$, $M_u = \bar{M} M_{\rm T}^{(u)}$, where $M_{\rm R}=\Theta(1)$ is the number of receive antennas, $M_{\rm T}^{(u)}=\Theta(1)$ is the number of transmit antennas of user $u$, and $\bar{M} = mn$ with $n$ blocks of $m$ chips each. $\mathbf{H}_u$ can be written as
\begin{align*}
    \mathbf{H}_u=\begin{bmatrix}
        \mathbf{H}^{1,1}_u&\cdots&\mathbf{H}^{1,M^{(u)}_{\rm T}}_u\\
        \vdots&\ddots&\vdots\\
        \mathbf{H}^{M_{\rm R},1}_u&\cdots&\mathbf{H}^{M_{\rm R},M^{(u)}_{\rm T}}_u
    \end{bmatrix},
\end{align*}
with $\mathbf{H}^{r,t}_u=\mathrm{blkdiag}\{\underbrace{\bar{\mathbf{H}}^{r,t}_u,\dots,\bar{\mathbf{H}}^{r,t}_u}_{n~\text{blocks}}\}$, $\bar{\mathbf{H}}^{r,t}_u\in\mathbb{C}^{m\times m}$ representing the static channel of one block, repeated across $n$ blocks.  The SR codes are assumed to satisfy Assumption~\ref{assumption: system scaling and power constraints}, which ensures that code rate $R_u = L_u \log B_u / \bar{M}$ (nats per received block symbol) converges to a finite constant and prevents any section from dominating or vanishing in power.
\begin{assumption}\label{assumption: system scaling and power constraints}
    Considering the regime where $m = \mathcal{O}(1)$ and $n, B_u, L_u \to \infty$, we assume that the aspect ratio $M_u / N_u$ converges to $\beta_u = \Theta(\log B_u / B_u)$, and that $p_{u,l} = \Theta(B_u)$.
\end{assumption}
\begin{remark}
    Above and throughout the sequel, for a sequence in $N\in\big\{M,\{N_u\}_{u=1}^U\big\}$ of random vectors $\mathbf a_N\in \mathbb C^{d}$ where $d$ is arbitrary, we write $\mathbf a_N=\mathcal O(\kappa)$ if $\left(\mathbb E [\Vert \mathbf a_N/\kappa \Vert^p]\right)^\frac {1}{p}\leq C_p$ for all $p\in \mathbb N$ and some constants $C_p$ independent of $N$. We further write $\mathbf a_N=\Theta(\kappa)$ if there exist constants $0 < c_p \leq C_p < \infty$ independent of $N$, such that $c_p \leq \left(\mathbb E [\Vert \mathbf a_N/\kappa \Vert^p]\right)^\frac {1}{p}\leq C_p$ holds for all $p\in \mathbb N$.
\end{remark}
% \begin{assumption}\label{assumption: matrices with bounded norm}
%     % With $m=\mathcal{O}(1),\ n\to\infty$, and an aspect ratio $\alpha_u\doteq M/M_u=M_{\rm R}/M^{(u)}_{\rm T}$, we assume that for $u=1,\dots,U$, $\Vert\bar{\mathbf{H}}^{r,t}_u\Vert_2=\mathcal{O}(1)$, where $\Vert\cdot\Vert_2 $ stands for the spectral norm of the matrix in the argument.
%     With $m=\mathcal{O}(1)$, we assume that $n\to\infty$.
% \end{assumption}

Moreover, under Assumption~\ref{assumption: system scaling and power constraints}, the capacity region\cite{Tse2005} of the considered MIMO-MAC can be characterized as follows. 
\begin{lemma}[MIMO-MAC Capacity Region]\label{lemma: capacity region}
The capacity region\cite{Tse2005} of the MIMO-MAC in~\eqref{equ: MIMO-MAC with mu-RT} is defined below for all $\mathcal{U} \subseteq [U]$, where $[U]\doteq\{1,\dots,U\}$:
    \begin{align}\label{equ: capacity region}
        \sum_{u\in\mathcal{U}}R_u&\leq \frac{1}{\bar{M}}\log{\Big\lvert\mathbf{I}+\mathrm{snr}\sum_{u\in\mathcal{U}}\mathbf{A}_u\mathbf{A}^\dagger_u\Big\rvert}=\mathcal O(1),
    \end{align}
where $\mathrm{snr} = \sigma^{-2}$, $R_u$ is the achievable rate per received block symbol of user $u$. Unless otherwise stated, all rate metrics in the sequel are measured in nats per received block symbol. 
\end{lemma}

Accordingly, the achievable sum rate is tightly upper bounded by the sum capacity $C_{\rm sum}$
\begin{align}\label{equ: sum capacity}
R_{\rm sum}&\doteq\sum_{u=1}^U R_u
\le C_{\rm sum}\doteq\frac{1}{\bar{M}}\log\Big\lvert \mathbf{I}
+\mathrm{snr}\mathbf{A}\mathbf{A}^\dagger \Big\rvert ,
\end{align}
where $\mathbf{A}\doteq[\mathbf{A}_1,\dots,\mathbf{A}_U]$.

\subsection{MA-OAMP Receiver}\label{subsec: MA-OAMP}
Multiple-access orthogonal approximate message passing (MA-OAMP) \cite{Burak2026} extends OAMP \cite{Ma2017} and VAMP \cite{Rangan2019vector} to multi-user scenarios, providing a robust multi-user signal detection technique. In contrast to the classical OAMP/VAMP, the state evolution (SE) of MA-OAMP forms a $U$-dim vector instead of a single scalar. 

Below, we collect the relevant findings of \cite{Burak2026}: for each user $u$, the algorithm begins with initializations $\mathbf{s}^{(1)}_u=\mathbf{0}$ and $ v^{(1)}_u=1$ and then proceeds for $t=1,\dots,T$ as follows:
\BS
\begin{align}
\mathbf{\Sigma}^{(t)}&=\big(\sigma^2\mathbf{I}+\sum_{u=1}^U{v^{(t)}_u}\mathbf{A}_u\mathbf{A}^\dagger_u\big)^{-1},\\
\mathbf{r}^{(t)}_u&=\mathbf{s}^{(t)}_u+\frac{1}{\chi^{(t)}_u}\mathbf{A}^\dagger_u\mathbf{\Sigma}^{(t)}\Big(\mathbf{y}-\sum_{i=1}^U\mathbf{A}_i\mathbf{s}^{(t)}_i\Big),\\
\bm{\eta}_u^{(t+1)}&= \mathbb E\big[\mathbf s_u\vert \mathbf r_u^{(t)}=\mathbf s_u+\sqrt{\tau_{u}^{(t)}}\mathbf z_u\big],
\\
\mathbf s_u^{(t+1)} &= \frac{\tau_u^{(t)}\bm{\eta}^{(t+1)}_u- \xi_u^{(t+1)} \mathbf{r}_u^{(t)}}{\tau_u^{(t)}-\xi_u^{(t+1)}},
\end{align} 
\ES
where each $\mathbf z_u\sim\mathcal{CN}(\mathbf 0,\mathbf I_{N_u})$ is an $N_u$-dim Gaussian random vector, and the necessary scalar parameters are constructed according to the SE recursion as
\BS
\label{SErec}
\begin{align}
\chi_u^{(t)} &= \frac{1}{N_u}{\rm tr} (\mathbf A_u^\dagger \mathbf\Sigma^{(t)} \mathbf A_u) ,\\
\tau_{u}^{(t)}&=\frac{1}{\chi_u^{(t)}}-v_{u}^{(t)},\\
\xi_u^{(t+1)}&=\frac{1}{N_u}\mathbb E~\Big[\big\Vert \mathbf s_u- \mathbb E\big[\mathbf s_u\vert \mathbf s_u+\sqrt{\tau_u^{(t)}}\mathbf z_u\big]\big\Vert^2\Big],\\
v_{u}^{(t+1)}&=\frac{\tau_{u}^{(t)}\xi_u^{(t+1)}}{\tau_{u}^{(t)}-\xi_u^{(t+1)}}.\label{equ: nonlinear SE}
\end{align}
\ES
Here, the SE recursion $\{\tau^{(t)}_u,v^{(t)}_u\}$ in~\eqref{SErec} is well-defined, and it has the following contraction property over the iteration steps:
\[
0 < v_u^{(t+1)} < v_u^{(t)} \quad \text{and} \quad 
0 < \tau_u^{(t+1)} < \tau_u^{(t)}.
\]

As established in \cite{Burak2026}, the proposed dynamics has the decoupling principle that for any $(u,t)\in [U]\times [T]$:
\begin{align}
\mathbf r_u^{(t)}=\mathbf s_u+\bm{\psi}_u^{(t)}+\mathcal O(1). \label{dpres}
\end{align}
For each $u\in[U]$, \(\{\bm{\psi}_u^{(t)} \in \mathbb C^{N_u}\}_{t}\) is an i.i.d. zero-mean Gaussian process that is independent of $\mathbf s_u$ such that
$\bm{\psi}_u^{(t)}\sim_{\text{i.i.d.}}\Psi_{u}^{(t)}$ with $\mathbb E\big[\vert\Psi_u^{(t)}\vert^2\big]=\tau_{u}^{(t)}$. We note that the result~\eqref{dpres} is non-asymptotic and it has the following asymptotic implication: for any small constant $\epsilon > 0$ 
\begin{equation}
\frac{\Vert \mathbf r_{u}^{(t)}-(\mathbf s_u+\bm{\psi}_{u}^{(t)}) \Vert }{N^{\epsilon}_u} \overset{\text{a.s.}}{\rightarrow} 0 \quad {\rm as}\quad N_u\to\infty.\label{ascon}
\end{equation} 

\section{Optimal Coding Principle for MA-OAMP Receiver}\label{sec: optimal coding principle for MIMO-MAC with MA-OAMP receiver}
% In this section, we show that the achievable user-rate suprema admit an integral representation; likewise, given that the rate-MMSE lemma\cite{Guo2005,Bhattad2007} also formulates the user rate as an integral over the same interval, we propose an optimal coding principle that forces the user rates to reach these suprema by matching the two integrands. Meanwhile, because the achievable user-rate supremum tuple lies on the dominant face of the MIMO-MAC capacity region and can achieve any point on this face when combined with time sharing \cite{Cover2006}, the resulting coding principle is capacity-region achieving.
This section develops a theoretical framework for optimal coding in MIMO-MAC with the MA-OAMP receiver. In Section~\ref{subsec: user-rate suprema}, we establish an analytical integral representation for the achievable user-rate suprema. Accordingly, in Section~\ref{subsec: achievable capacity region}, we propose an optimal coding principle under which the code rates asymptotically attain these suprema. Moreover, we show that codes designed under this principle can achieve the entire capacity region when combined with time sharing \cite{Cover2006}.
\subsection{Achievable User-Rate Suprema}\label{subsec: user-rate suprema}
MA-OAMP involves multiple variance variables, and we adopt the following componentwise monotonically decreasing assumption on the variances in~\eqref{equ: nonlinear SE}, with the iteration index $(t)$ dropped for brevity, as it applies within the same iteration.

\begin{assumption}\label{assumption: componentwise monotonically decreasing variance path}
Let $\mathbf{v}(\theta) \doteq[v_1(\theta),\dots,v_U(\theta)]^T$ denote the variances in~\eqref{equ: nonlinear SE}, implicitly parameterized by $\theta \in [0,1]$, with $\mathbf{v}(0)=\mathbf{1}$ and $\mathbf{v}(1)=\mathbf{0}$, satisfying: 1) $\mathbf{v}(\theta_1)\neq\mathbf{v}(\theta_2)\implies\mathbf{v}(\theta_1)\succeq\mathbf{v}(\theta_2)$ or $\mathbf{v}(\theta_1)\preceq\mathbf{v}(\theta_2)$; 2) $\Vert\mathbf{v}(\theta_1)\Vert_1>\Vert\mathbf{v}(\theta_2)\Vert_1\implies0\le\theta_1<\theta_2\le1$.
\end{assumption}

In the sequel, the variances $v_u,~\forall~u\leq U
$ are assumed to satisfy Assumption~\ref{assumption: componentwise monotonically decreasing variance path}, with $\theta$ omitted. Under Assumption~\ref{assumption: componentwise monotonically decreasing variance path}, the following lemma gives the achievable user-rate suprema.
\begin{lemma}[Achievable User-Rate Suprema]\label{lemma: achievable user-rate suprema under MA-OAMP}
For user $u$, the achievable rate of MA-OAMP at the limit signal-to-noise ratio (SNR) $\mathrm{snr}=\sigma^{-2}$ is tightly upper bounded by $R^{\rm sup}_u$, i.e., $R_u\le R^{\rm sup}_u$, with
\BS\label{equ: achievable user rate supremum}
\begin{align}
R^{\rm sup}_u &= \frac{N_u}{\bar{M}}\int_0^1\min\left\{\mathcal{F}_u(\xi),\hat{\phi}^{-1}_{\rm Gau}(\xi)\right\}\mathrm{d}\xi\label{equ: achievable user rate supremum with VTF}\\
&\equiv\frac{1}{\bar{M}}\! \int_0^1\!\! \mathrm{tr}\left\{\!\mathbf{A}^\dagger_u\Big(\sigma^2\mathbf{I}+ \!\! \sum_{i=1}^Uv_i\mathbf{A}_i\mathbf{A}^\dagger_i\Big)^{\!-1}\mathbf{A}_u\!\right\}\mathrm{d}v_u,\label{equ: analytical user rate supremum}   
\end{align}
Here, $\hat{\phi}^{-1}_{\rm Gau}(\cdot)$ denotes the functional inverse of the Gaussian MMSE $\hat{\phi}_{\rm Gau}(\varrho)=(1+\varrho)^{-1}$, and the variational transfer function (VTF) for each $u=1,\dots,U$ is defined as
\begin{align}\label{equ: VTF}
\mathcal F_u(\xi)&\doteq \xi^{-1}-{S^{-1}_u(\xi)}, 
\end{align}    
with $S^{-1}_u(\cdot)$ denoting the functional inverse of

\begin{align}\label{equ: S_u}
    S_u(\rho)&\doteq\frac{1}{N_u}\mathrm{tr}\left\{\big(\mathbf{A}^\dagger_u\mathbf{\Sigma}_{\backslash u}\mathbf{A}_u+\rho\mathbf{I}\big)^{-1}\right\},~\rho \in (0, \infty),
\end{align}
\ES
where $\mathbf{\Sigma}_{\backslash u}=\big(\sigma^2\mathbf{I} + \sum_{i\neq u} v_i \mathbf{A}_i \mathbf{A}^\dagger_i\big)^{-1}$. 
\end{lemma}

\begin{remark}
    Formally, the limit SNR, denoted as $\mathrm{snr} = \sigma^{-2}$, is defined for a given deterministic sum capacity $C_{\rm sum}=\frac{1}{\bar{M}}\log{|\mathbf{I}+\mathrm{snr}\mathbf{A}\mathbf{A}^\dagger|}$. 
\end{remark}
\begin{remark}
    The integral in~\eqref{equ: analytical user rate supremum} is a path integral that depends on the variance path $\mathbf{v}=[v_1,\dots,v_u,\cdots,v_U]^T$. Under Assumption~\ref{assumption: componentwise monotonically decreasing variance path}, the remaining components $v_i~(i\neq u)$ monotonically decrease as a function of the implicit parameter $\rho~(\doteq v^{-1}_u)$, i.e., $v_i:=v_i(\rho),~i\neq u$, which ensures the existence of $S^{-1}_u(\cdot)$ in~\eqref{equ: VTF}. For clarity, the implicit parameter $\rho$ is omitted for all variances in~\eqref{equ: analytical user rate supremum} and~\eqref{equ: S_u}.
\end{remark}
\begin{figure}[!htbp]
    \centering
    \includegraphics[width=0.7\linewidth]{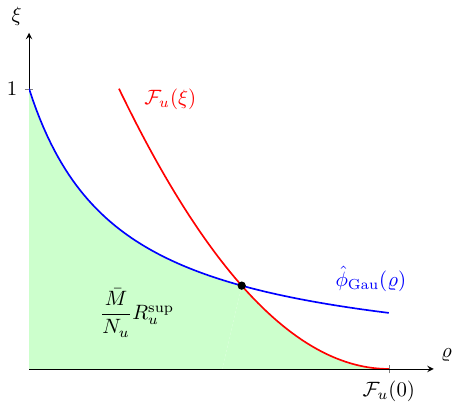}
    \caption{Achievable rate supremum $R^{\rm sup}_u$ for user $u$ in Lemma~\ref{lemma: achievable user-rate suprema under MA-OAMP}.}
    \label{fig: user-rate suprema}
\end{figure}

As illustrated in Fig.~\ref{fig: user-rate suprema}, Lemma~\ref{lemma: achievable user-rate suprema under MA-OAMP} shows that the achievable user-rate suprema of the MA-OAMP receiver correspond to the area in the positive quadrant under the curve defined by the minimum of: (i) the MMSE function for Gaussian signals, and (ii) the VTF of MA-OAMP. Due to page limits, proofs of key lemmas, corollaries, and theorems are deferred to the extended version \cite{Hao2026}. We thus briefly explain the intuition behind Lemma~\ref{lemma: achievable user-rate suprema under MA-OAMP}:

\emph{1)}~The minimum in~\eqref{equ: achievable user rate supremum with VTF} stems from two facts: first, as shown in the following Lemma~\ref{lemma: replica mmse in multi-user systems}, the achievable MSE of the MA-OAMP receiver corresponds to the solution of the fixed-point equation $\mathcal{F}_u(\xi)=\hat{\phi}^{-1}_u(\xi)$, where $\hat{\phi}^{-1}_u(\cdot)$ is the functional inverse of the MMSE function $\hat{\phi}_u(\cdot)$ of signal $\mathbf{s}_u$: 
\begin{equation}\label{equ: MMSE}
 \hat{\phi}_u(\varrho)\doteq \frac{1}{N_u}\mathbb E~\left[\big\Vert \mathbf s_u- \mathbb E\big[\mathbf s_u\vert \sqrt{\varrho}\mathbf s_u+\mathbf z\big]\big\Vert^2\right],~ \mathbf{z}\sim\mathcal{CN}(\mathbf{0},\mathbf{I}).
\end{equation}
\begin{lemma}[Achievable MSE]\label{lemma: replica mmse in multi-user systems}
The achievable MSE of user $u$, denoted by $\xi^*_u$, is the maximum fixed point of
\begin{align}\label{equ: fixed-point equation}
    \hat{\phi}^{-1}_u\left(\xi\right)=\mathcal{F}_u\left(\xi\right).
\end{align}
\end{lemma}
Thus, asymptotically error-free detection requires the MMSE function of the signal to avoid fixed points with the VTF $\mathcal{F}_u(\xi)$; second, the MMSE function of any signal cannot exceed that of an i.i.d. Gaussian signal.

\emph{2)}~The integral arises from the following rate-MMSE lemma (Lemma~\ref{lemma: Rate-mmse}), which establishes a relationship between the MMSE function and the rate of the signal. 
% Due to page limits, proofs of key lemmas, corollaries, and theorems are deferred to the extended version \cite{Hao2026}.
% Likewise, the rate-MMSE lemma \cite{Guo2005,Bhattad2007} relates the rate of a code to the area under its MMSE function.
\begin{lemma}[Rate-MMSE\cite{Guo2005,Bhattad2007}]\label{lemma: Rate-mmse}
The rate of signal $\mathbf{s}_u$ is
\begin{align}
    R_u=\frac{N_u}{\bar{M}}\int_0^\infty\hat{\phi}_u(\varrho)\mathrm{d}\varrho=\frac{N_u}{\bar{M}}\int_0^1\hat{\phi}^{-1}_u(\xi)\mathrm{d}\xi.\label{equ: rate-mmse}
\end{align}
\end{lemma} 
\subsection{Capacity-Region-Achieving Optimal Coding Principle}\label{subsec: achievable capacity region}
By combining Lemma~\ref{lemma: achievable user-rate suprema under MA-OAMP} and Lemma~\ref{lemma: Rate-mmse}, signals that achieve the suprema in~\eqref{equ: achievable user rate supremum} can be designed by matching $\hat{\phi}_u^{-1}(\xi)$ to the minimum of (i) the MMSE function for Gaussian signals, and (ii) the VTF, i.e., $\hat{\phi}^{-1}_u(\xi)\approx \varphi_u(\xi)$, where $
\varphi_u(\xi)\doteq\min\left\{\mathcal{F}_u(\xi),\hat{\phi}^{-1}_{\rm Gau}(\xi)\right\}$. Furthermore, as shown in Lemma~\ref{lemma: replica mmse in multi-user systems}, it suffices to design the MMSE function such that 
for any $\xi \in \left(0,1\right)$, $\hat{\phi}^{-1}_u(\xi) < \varphi_u(\xi)$, which prevents any fixed point between $\hat{\phi}_u(\xi)$ and $\mathcal{F}_u(\xi)$ and thus guarantees asymptotically vanishing error at the limit SNR. Consequently, codes with vanishing probability of error achieving the user-rate suprema in~\eqref{equ: achievable user rate supremum} are theoretically attainable if:

\BS
\begin{enumerate}
    \item $\hat{\phi}^{-1}_u(\cdot)$ has no fixed point with $\mathcal{F}_u(\cdot)$, that is
\begin{align}
    \hat{\phi}^{-1}_u(\xi)\lesssim\varphi_u(\xi)<\mathcal{F}_u(0),~\forall~\xi\in\left(0,1\right).
\end{align}
    \item $R_u$ approaches the supremum $R^{\rm sup}_u$, that is
\begin{align}
    \frac{N_u}{\bar{M}}\int_0^1\hat{\phi}^{-1}_u(\xi)\mathrm{d}\xi\lesssim\frac{N_u}{\bar{M}}\int_0^1\varphi_u(\xi)\mathrm{d}\xi.
\end{align}
\end{enumerate}
\ES
Accordingly, we introduce the following Assumption~\ref{assumption: optimal coding principle}.
\begin{assumption}[Optimal Coding Principle]\label{assumption: optimal coding principle} 
For each user $u$, there exists a signal distribution for $\mathbf{s}_u$ such that the MMSE function $\hat{\phi}_u(\cdot)$ satisfies
    \begin{align}\label{equ: optimal coding principle}
        0<\varphi_u(\xi)-\hat{\phi}^{-1}_u(\xi)=\mathrm{o}(\beta_u),~\forall\ \xi \in (0,1).
    \end{align}  
\end{assumption}
Under Assumption~\ref{assumption: optimal coding principle}, user rates achieve the suprema if the signal MMSE functions satisfy~\eqref{equ: optimal coding principle}, because \[
R^{\rm sup}_u-R_u= \frac{N_u}{\bar{M}}\int_0^1\big[\varphi_u(\xi)-\hat{\phi}^{-1}_u(\xi)\big]\mathrm{d}\xi=\mathrm{o}(1).
\]
\begin{remark}
    In this section, we only assume the existence of such an MMSE function. Explicit constructions of SR codes satisfying~\eqref{equ: optimal coding principle} will be presented in Section~\ref{sec: capacity-region-achieving SR codes via power allocation}.
\end{remark}
% The specific construction of SR codes meeting~\eqref{equ: optimal coding principle} is deferred to Section~\ref{sec: capacity-region-achieving SR codes via power allocation}.

Meanwhile, it can be shown that signal designs complying with~\eqref{equ: optimal coding principle} enable user rates to attain 1) the sum capacity, 2) the vertices of the dominant face, and 3) the entire MIMO-MAC capacity region:
\subsubsection{Achieving Sum Capacity}
The following Corollary~\ref{corollary: sum rate supremum} shows that the achievable sum rate of the MA-OAMP receiver attains the sum capacity of the MIMO-MAC and hence lies on the dominant face of the capacity region, provided that the MMSE functions satisfy~\eqref{equ: optimal coding principle}.
\begin{corollary}\label{corollary: sum rate supremum}
Given the MMSE functions $\{\hat{\phi}_u\}_{u=1}^U$ that satisfy~\eqref{equ: optimal coding principle}, the achievable sum rate of the MA-OAMP receiver achieves the sum capacity of the MIMO-MAC, namely, $0<\sum_{u=1}^UR^{\rm sup}_u-R_{\rm sum}=\mathrm{o}(1)$, where $\sum_{u=1}^UR^{\rm sup}_u\equiv C_{\rm sum}$.
\end{corollary}

\subsubsection{Achieving Vertices of the Dominant Face}
Let $\mathcal{P}_{\rm perm}$ denote the set of all permutations of $\{1,\dots,U\}$. For each permutation $\mathcal{P}^j_{\rm perm}=[k^j_1,\dots,k^j_U]\in\mathcal{P}_{\rm perm}$, define
\[
\mathcal{K}^j_i\doteq[k^j_1,\dots,k^j_i],~
\mathbf{A}_{\mathcal{K}^j_i}\doteq \big[\mathbf{A}_{k^j_1},\dots,\mathbf{A}_{k^j_i}\big].
\]
Then, the corresponding vertex $\mathcal{C}^{\rm vert}_j=(C^{\rm vert}_1,\dots,C^{\rm vert}_U)$ of the dominant face\cite{Tse2005} is
\BS\label{equ: vertex j}
\begin{align}
C^{\rm vert}_{k^j_1}=&\frac{1}{\bar{M}}\log\bigl|\mathbf{I}+\mathrm{snr}\mathbf{A}_{k^j_1}\mathbf{A}^\dagger_{k^j_1}\bigr|,\\
C^{\rm vert}_{k^j_2}=&\frac{1}{\bar{M}}\log\frac{\bigl|\mathbf{I}+\mathrm{snr}\mathbf{A}_{\mathcal{K}^j_2}\mathbf{A}^\dagger_{\mathcal{K}^j_2}\bigr|}{\bigl|\mathbf{I}+\mathrm{snr}\mathbf{A}_{k^j_1}\mathbf{A}^\dagger_{k^j_1}\bigr|},
\\\vdots\nonumber\\
C^{\rm vert}_{k^j_U}=&\frac{1}{\bar{M}}\log\frac{\bigl|\mathbf{I}+\mathrm{snr}\mathbf{A}\mathbf{A}^\dagger\bigr|}{\bigl|\mathbf{I}+\mathrm{snr}\mathbf{A}_{\mathcal{K}^j_{U-1}}\mathbf{A}^\dagger_{\mathcal{K}^j_{U-1}}\bigr|}.
\end{align}
\ES
The following Corollary~\ref{corollary: achievable vertices} states that under Assumption~\ref{assumption: optimal coding principle}, the MA-OAMP receiver is capable of achieving these vertices by choosing an SIC variance path.
\begin{corollary}\label{corollary: achievable vertices}
Given the MMSE functions $\{\hat{\phi}_u\}_{u=1}^U$ that satisfy~\eqref{equ: optimal coding principle}, consider an SIC variance path defined by\cite{Liu2019}:
\begin{align}
    \gamma_{k^j_{i+1}}(v^{-1}_{k^j_{i+1}}-1)=\gamma_{k^j_i}(v^{-1}_{k^j_i}-1),~\forall~i=1,\dots,U-1.\label{equ: path j}
\end{align}
If $\kappa_i=\gamma_{k^j_i}/\gamma_{k^j_{i+1}}\to\infty$, the MA-OAMP receiver can achieve the vertex in~\eqref{equ: vertex j}, i.e., $R_{k^j_i}\to C^{\rm vert}_{k^j_i},~\forall i=1,\dots,U$.
\end{corollary}

\subsubsection{Achieving the Entire Capacity Region}
 As the MIMO-MAC capacity region is a convex polytope with all interior points dominated by its dominant face, any rate tuple on this face can be achieved via convex combinations of its vertices using time sharing \cite{Cover2006}. Since Corollary~\ref{corollary: achievable vertices} establishes that all vertices of the dominant face are attainable with a properly designed SIC path, we conclude that the entire MIMO-MAC capacity region is achievable under Assumption~\ref{assumption: optimal coding principle}.
% \begin{corollary}\label{corollary: achievable capacity region}
% Given the MMSE functions $\{\hat{\phi}_u\}_{u=1}^U$ that satisfy~\eqref{equ: optimal coding principle}, the MA-OAMP receiver is capable of achieving the entire capacity region of MIMO-MAC, along with the time-sharing technique\cite{Cover2006}. 
% \end{corollary}

\section{Capacity-Region-Achieving SR Codes via Power Allocation}\label{sec: capacity-region-achieving SR codes via power allocation}
We analyze the MMSE function of the position-modulated signal, which forms the basis for designing capacity-region-achieving SR codes following the optimal coding principle~\eqref{equ: optimal coding principle}. For user $u$ with power allocation $\mathbf{p}_u\doteq[p_{u,1},\dots,p_{u,L_u}]^T$, the MMSE exhibits phase transitions characterized by the following Lemma~\ref{lemma: asymptotic MMSE} in the asymptotic regime.
\begin{lemma}[Asymptotic MMSE of Position-Modulated Signals]\label{lemma: asymptotic MMSE}
Under Assumption~\ref{assumption: system scaling and power constraints}, the MMSE function satisfies
\begin{align}
\hat{\phi}_u(\varrho) 
&\simeq  \sum_{l=1}^{L_u} P_{u,l} \hat{\phi}_{\rm Gau}(\varrho P_{u,l})\cdot\mathbb{I}\Big(\frac{\log {B_u}}{p_{u,l-1}}\leq \varrho<\frac{\log {B_u}}{p_{u,l}}\Big),\label{equ: asymptotic MMSE between different phase transition points}
\end{align}
where $\log {B_u}/p_{u,0} \doteq 0$, and $P_{u,l} \doteq \sum_{k=l}^{L_u} p_{u,k}/N_u$ is the residual power. $\mathbb{I}(\cdot)$ denotes the indicator function.
\end{lemma}
\begin{remark}
    Notably, Lemma~\ref{lemma: asymptotic MMSE} is asymptotically equivalent to \cite[Lemma~1]{Rush2017} under the typical regime $1/p_{u,1}=\mathrm{o}(1/\log{B_u})$. However, directly applying \cite[Lemma~1]{Rush2017} yields insufficiently accurate MMSE function approximations, which prevents us from satisfying~\eqref{equ: optimal coding principle}. This constitutes the key motivation for developing a more precise asymptotic MMSE.
% Compared to \cite[Lemma 1]{Rush2017}, Lemma~\ref{lemma: asymptotic MMSE} provides a more precise characterization of the phase transitions, aiming to meet the optimal coding principle~\eqref{equ: optimal coding principle}.
\end{remark}
\subsection{Capacity-Region-Achieving Power Allocation}
 Given Lemma~\ref{lemma: asymptotic MMSE}, the following Theorem~\ref{theorem: existence of a capacity-achieving src 2} shows that, for a given target rate suprema tuple $\mathcal{R}^{\rm sup}\doteq(R^{\rm sup}_1,\dots,R^{\rm sup}_U)$, with a corresponding variance path satisfying Assumption~\ref{assumption: componentwise monotonically decreasing variance path}, SR codes can be constructed via an iterative power allocation scheme that fulfills the optimal coding principle in~\eqref{equ: optimal coding principle}, thereby achieving $\mathcal{R}^{\rm sup}$ on the dominant face. Notably, as proven in the extended version \cite{Hao2026}, this power allocation reduces to that of \cite[Theorem 2]{Xu2023} for $U=1$, and further matches the exponentially decaying power allocation proposed for AWGN channels in \cite{Rush2017}.

\begin{theorem}[Capacity-Region-Achieving SR Codes]\label{theorem: existence of a capacity-achieving src 2}
Under Assumption~\ref{assumption: system scaling and power constraints}, consider a target rate tuple $\mathcal{R}^{\rm sup}=(R^{\rm sup}_1,\dots,R^{\rm sup}_U)$ on the dominant face, together with a corresponding variance path that satisfies Assumption~\ref{assumption: componentwise monotonically decreasing variance path}. Then, by initializing $k=1$ and $P_{u,1}=1$, a capacity-region-achieving power allocation vector $\mathbf{p}_u$ can be recursively obtained as:

\begin{enumerate}
\BS\label{equ: PA_SRC}

    \item Find the solution $\varrho^*_{u,k}$ to the following equation:
    \begin{align}
        \mathcal{F}^{-1}_u(\varrho)=P_{u,k}\hat{\phi}_{\rm Gau}(\varrho P_{u,k}).
    \end{align}
    \item Compute:
    \begin{align}
    &p_{u,k} = \frac{\log B_u}{\varrho^*_{u,k}}+\epsilon_k,~\epsilon_k=\mathrm{o}\Big(\frac{\log{B_u}}{\varrho^*_{u,k}}\Big)>0,\\
    &P_{u,k+1} = P_{u,k} - \frac{p_{u,k}}{N_u},
    \end{align}
\ES
\end{enumerate}
\end{theorem}
% \begin{remark}\label{remark: iter PA is equiv to exp PA}
%     For the case of $U=1$, the power allocation in~\eqref{equ: PA_SRC} coincides with \cite[Theorem~2]{Xu2023}, yielding a simple and practical solution that extends it naturally to MIMO-MAC. For single-user AWGN channels, it further reduces to the exponentially decaying power allocation\cite{Rush2017}\cite{Xu2023}.
% \end{remark}

\subsection{Improved Power Allocation for Finite-Length SR Codes}
Theorem~\ref{theorem: existence of a capacity-achieving src 2} provides a power allocation under asymptotic conditions. However, with finite channel dimensions and section lengths, SR codes incur performance loss, motivating optimized power allocation to improve practical performance. Lemma~\ref{lemma: replica mmse in multi-user systems} and the following Lemma~\ref{lemma: replica map ser in multi-user systems}\cite[Proposition~1]{Greig2018} together characterize the achievable maximum a posteriori (MAP) section error rate (SER) of MA-OAMP.

\begin{lemma}[Achievable MAP SER\cite{Greig2018}]\label{lemma: replica map ser in multi-user systems}
Given achievable MSE $\xi^*_u$, the converged input SNR for decoder is $\varrho^*_u=\hat{\phi}^{-1}_u\left(\xi^*_u\right)$. Accordingly, the MAP SER of user $u$ for the observation $\mathbf{y}=\sqrt{\varrho^*_u}\mathbf{s}_u+\mathbf{z}_u$ is 
\begin{align}\label{equ: map ser}
    \mathcal{Q}_u(\varrho^*_u)
    =1-\frac{1}{L_u}\sum_{l=1}^{L_u}\mathbb{E}_z\left[\Phi\Big(z+\sqrt{2\varrho^*_up_{u,l}}\Big)\right]^{B_u-1},
\end{align}
where $z\sim\mathcal{N}(0,1)$ and $\Phi(\cdot)$ is the cumulative distribution function (CDF) of the standard normal distribution. Note that $\xi^*_u$ is determined by the power allocation scheme $\mathbf{p}_u$ over the position-modulated signal.
\end{lemma}
% \begin{figure}[!t]
%     \centering
%     \includegraphics[width=0.8\linewidth]{figures/pngfig/result,U=1/result,U=1,M=2^13,v_n=0.2,DS=1.04e-7,v=0,AWGN,target=2.png}
%         \label{fig: results,U=1,AWGN,high rate}
%     \caption{BER performance in a single-user AWGN channel. The noise power $\sigma^2=0.2$. The target rate is $R^{\rm sup}_1=2.585$~bits/sym. The SR code parameters are $(B,L)=(1939,1939)$. 
% The code length is $8192$. The underlying LDPC code has rate $R_{\rm LDPC}=0.646$~bits/sym and length 32768, yielding a symbol sequence of length 8192 and a spectral efficiency of $R^{\rm sup}_1$ after 16-QAM modulation.}
%     \label{fig: results,U=1,AWGN}
% \end{figure}

Given the section length $B_u$, the number of sections $L_u$, and channel state information (CSI), decoding can be optimized by minimizing the MAP SER in Lemma~\ref{lemma: replica map ser in multi-user systems}, i.e.,
\BS\label{equ: optimization problem}
\begin{align}
    &\underset{\mathbf{p}_u}{\min}~\mathcal{Q}_u\left(\varrho^*_u\right),\\
    &\mathrm{s.t.}~\sum_{l=1}^{L_u}p_{u,l}=N_u,\\
    &\qquad p_{u,1}\geq p_{u,2}\ge\cdots \geq p_{u,L_u}.
\end{align}
\ES

% To reduce repeated computation, we construct a lookup table of the single-section MMSE
% \[
%         \bar{\phi}_u(\varrho)=\frac{1}{B_u}\mathbb{E}\left[\frac{\sum_{i=2}^{B_u}e^{\sqrt{2\varrho}z_{l,i}}}{e^{\sqrt{2\varrho}z_{l,1}+2\varrho}+\sum_{i=2}^{B_u}e^{\sqrt{2\varrho}z_{l,i}}}\right],
%     \]
% where for any $i=1,\dots,B_u$, $z_{l,i}\sim\mathcal{N}(0,1)$. Then, given arbitrary $\mathbf{p}_u\in\mathbb{R}^{L_u}$, the MMSE is efficiently approximated as
%     \[
%         \hat{\phi}_u(\varrho)=\frac{1}{L_u}\sum_{l=1}^{L_u}p_{u,l}\bar{\phi}_u(\varrho p_{u,l}).
%     \]
Optimization is performed at a target SNR slightly above the limit SNR to ensure reliable decoding. For example, in an AWGN channel with $C_{\rm sum}=1$ bit (limit SNR $\mathrm{snr}=0~\mathrm{dB}$), we can set the target noise power to $\sigma^2_{\rm tar}\approx0.5$, corresponding to a target SNR of 3 dB. Empirically, under our settings, the best performance is achieved when the optimization is conducted at a target SNR about 1-3 dB above the limit SNR.

\section{Numerical Results}
Channels follow a Tapped Delay Line Type A (TDL-A) model with $K=23$ paths. Each transmit antenna uses $n=32$ blocks, each containing $m=256$ subcarriers with bandwidth $W=40$ MHz and carrier frequency $f_c=4$ GHz. Distinct channel realizations across users are ensured by assigning different power gains and random phases per user. The path in~\eqref{equ: path j} is used to align the achievable user-rate suprema with target $\mathcal{R}^{\rm sup}$ by adjusting coefficients $\{\gamma_u\}_{u=1}^U$ accordingly. We evaluate SR codes under four power allocation schemes: the iterative power allocation in Theorem~\ref{theorem: existence of a capacity-achieving src 2} (iterative PA), the optimized power allocation obtained by solving~\eqref{equ: optimization problem} (optimized PA), the average power allocation (average PA), and the exponentially decaying power allocation\cite{Rush2017} (exponential PA). 5G-NR LDPC codes serve as a benchmark, where n-QAM modulation is used to match the target spectral efficiency. SR codes are decoded using up to 100 MA-OAMP iterations, whereas 5G-NR LDPC decoding adopts a 1-step LMMSE equalizer followed by 100 belief propagation (BP) iterations. Rates are measured in bits per received block symbol (bits/sym), and SER for SR codes is converted to the bit error rate (BER) for a fair comparison. For the two-user MIMO-MAC with asymmetric CSI (Fig.~\ref{fig: results,U=2}), only iterative PA and optimized PA enable successful decoding; in particular, optimized PA achieves a BER below $10^{-6}$ within $2.8~\mathrm{dB}$ of the limit SNR, whereas all other coding schemes fail to decode.
\begin{figure}[!htbp]
    \centering

    \includegraphics[width=0.8\linewidth]{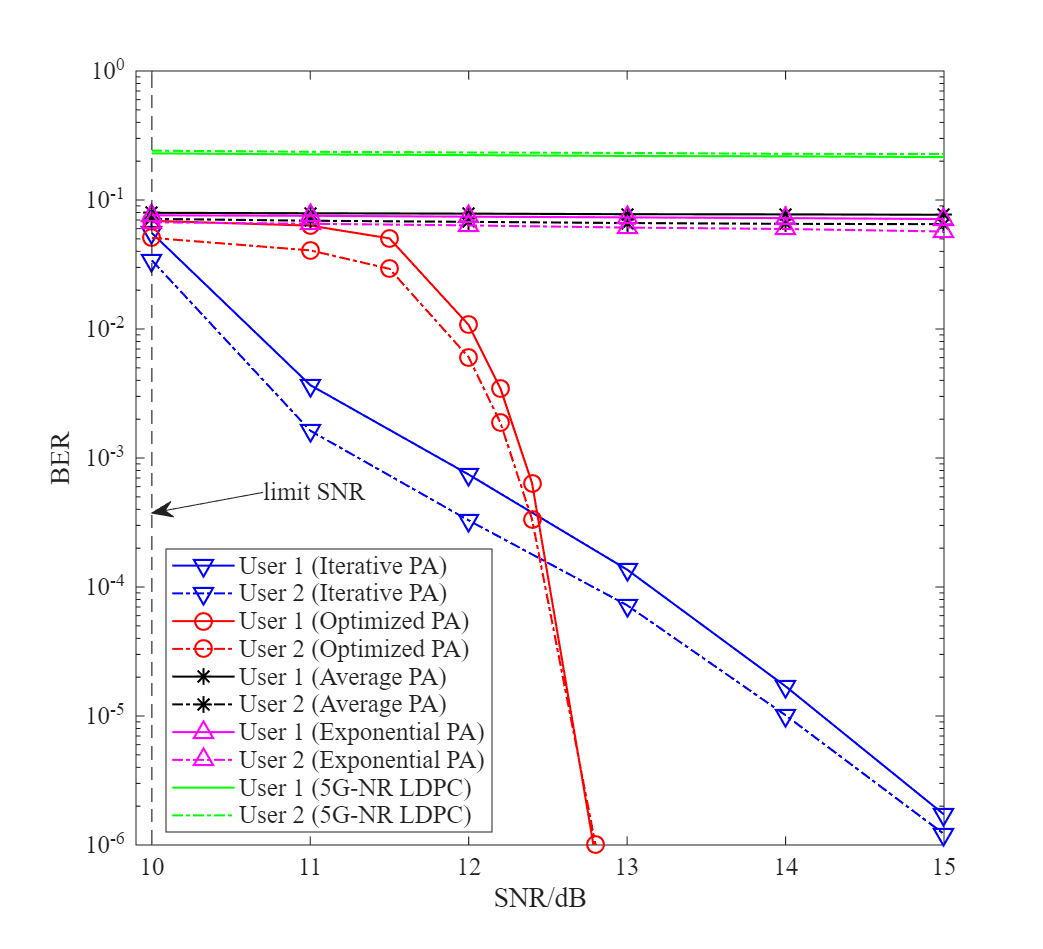}
       
    \caption{BERs in a two-user $2\times2$ MIMO-MAC. The limit SNR is $10$~dB. The randomly selected target rate tuple is $\mathcal{R}^{\rm sup}=(4.704,2.354)$~bits/sym. The power gains are $(0,-3)$~dB. The SR code parameters are $(B_1,L_1)=(1783,1783)$ and $(B_2,L_2)=(971,971)$. The underlying LDPC codes have rates $\mathcal{R}_{\rm LDPC}=(1.1760,1.1768)$~bits/sym and lengths (32768,16384), yielding symbol sequence lengths (8192,8192) and spectral efficiencies $\mathcal{R}^{\rm sup}$ after 16-QAM (User 1) and QPSK (User 2) modulation.}
    \label{fig: results,U=2}
\end{figure}
\section{Conclusion}

This paper proposes a capacity-region-achieving coding framework for MIMO-MAC using SR codes with the MA-OAMP receiver. Guided by an optimal coding principle that enables the MA-OAMP receiver to achieve the entire capacity region, capacity-region-achieving SR codes are designed via proper power allocation over position-modulated signals. Numerical results demonstrate that SR codes with the proposed power allocations outperform those with conventional power allocation schemes, as well as 5G-NR LDPC codes, exhibiting near capacity-region-achieving performance in MIMO-MAC.

%%%%%%
%% References:
%% We recommend the usage of BibTeX:
%%
\clearpage
\bibliographystyle{IEEEtran}
\bibliography{reference}

@INPROCEEDINGS{Berrou1993Turbo,
  author={Berrou, C. and Glavieux, A. and Thitimajshima, P.},
  booktitle={Proceedings of ICC '93 - IEEE International Conference on Communications}, 
  title={Near {Shannon} limit error-correcting coding and decoding: Turbo-codes. 1}, 
  year={1993},
  volume={2},
  number={},
  pages={1064-1070 vol.2},
  doi={10.1109/ICC.1993.397441}}

@ARTICLE{Richardson2001,
  author={Richardson, T.J. and Urbanke, R.L.},
  journal={IEEE Transactions on Information Theory}, 
  title={The capacity of low-density parity-check codes under message-passing decoding}, 
  year={2001},
  volume={47},
  number={2},
  pages={599-618},
  doi={10.1109/18.910577}}

@ARTICLE{Kudekar2011,
  author={Kudekar, Shrinivas and Richardson, Thomas J. and Urbanke, Rüdiger L.},
  journal={IEEE Transactions on Information Theory}, 
  title={Threshold Saturation via Spatial Coupling: Why Convolutional {LDPC} Ensembles Perform So Well over the {BEC}}, 
  year={2011},
  volume={57},
  number={2},
  pages={803-834},
  doi={10.1109/TIT.2010.2095072}}

@ARTICLE{Arikan2009,
  author={Arikan, Erdal},
  journal={IEEE Transactions on Information Theory}, 
  title={Channel Polarization: A Method for Constructing Capacity-Achieving Codes for Symmetric Binary-Input Memoryless Channels}, 
  year={2009},
  volume={55},
  number={7},
  pages={3051-3073},
  doi={10.1109/TIT.2009.2021379}}

@ARTICLE{Joseph2012,
  author={Joseph, Antony and Barron, Andrew R},
  journal={IEEE Transactions on Information Theory}, 
  title={Least Squares Superposition Codes of Moderate Dictionary Size Are Reliable at Rates up to Capacity}, 
  year={2012},
  volume={58},
  number={5},
  pages={2541-2557},
  doi={10.1109/TIT.2012.2184847}}

@ARTICLE{Joseph2014,
  author={Joseph, Antony and Barron, Andrew R.},
  journal={IEEE Transactions on Information Theory}, 
  title={Fast Sparse Superposition Codes Have Near Exponential Error Probability for ${R}<\mathcal{C}$}, 
  year={2014},
  volume={60},
  number={2},
  pages={919-942},
  doi={10.1109/TIT.2013.2289865}}

@ARTICLE{Rush2017,
  author={Rush, Cynthia and Greig, Adam and Venkataramanan, Ramji},
  journal={IEEE Transactions on Information Theory}, 
  title={Capacity-Achieving Sparse Superposition Codes via Approximate Message Passing Decoding}, 
  year={2017},
  volume={63},
  number={3},
  pages={1476-1500},
  doi={10.1109/TIT.2017.2649460}}

@article{Xu2023,
  title={Capacity-Achieving Sparse Regression Codes via Vector Approximate Message Passing},
  author={Yizhou Xu and Yuhao Liu and Shansuo Liang and Tingyi Wu and Bo Bai and Jean Barbier and Tianqi Hou},
  journal={2023 IEEE International Symposium on Information Theory (ISIT)},
  year={2023},
  pages={785-790}}

@ARTICLE{Liu2021,
  author={Liu, Lei and Liang, Chulong and Ma, Junjie and Ping, Li},
  journal={IEEE Transactions on Information Theory}, 
  title={Capacity Optimality of {AMP} in Coded Systems}, 
  year={2021},
  volume={67},
  number={7},
  pages={4429-4445},
  doi={10.1109/TIT.2021.3083748}}

@ARTICLE{Liu2024,
  author={Liu, Lei and Liang, Shansuo and Ping, Li},
  journal={IEEE Transactions on Communications}, 
  title={On Capacity Optimality of {OAMP}: Beyond IID Sensing Matrices and Gaussian Signaling}, 
  year={2024},
  volume={72},
  number={5},
  pages={2519-2535},
  doi={10.1109/TCOMM.2024.3354201}}

@ARTICLE{Ma2017,
  author={Ma, Junjie and Ping, Li},
  journal={IEEE Access}, 
  title={Orthogonal {AMP}}, 
  year={2017},
  volume={5},
  number={},
  pages={2020-2033},
  doi={10.1109/ACCESS.2017.2653119}}

@ARTICLE{Guo2005,
  author={Dongning Guo and Shamai, S. and Verdu, S.},
  journal={IEEE Transactions on Information Theory}, 
  title={Mutual information and minimum mean-square error in Gaussian channels}, 
  year={2005},
  volume={51},
  number={4},
  pages={1261-1282},
  doi={10.1109/TIT.2005.844072}}

@ARTICLE{Rush2019,
  author={Rush, Cynthia and Venkataramanan, Ramji},
  journal={IEEE Transactions on Information Theory}, 
  title={The Error Probability of Sparse Superposition Codes With Approximate Message Passing Decoding}, 
  year={2019},
  volume={65},
  number={5},
  pages={3278-3303},
  doi={10.1109/TIT.2018.2882177}}

@INPROCEEDINGS{Hou2022,
  author={Hou, TianQi and Liu, YuHao and Fu, Teng and Barbier, Jean},
  booktitle={2022 IEEE International Symposium on Information Theory (ISIT)}, 
  title={Sparse superposition codes under {VAMP} decoding with generic rotational invariant coding matrices}, 
  year={2022},
  volume={},
  number={},
  pages={1372-1377},
  doi={10.1109/ISIT50566.2022.9834843}}

@ARTICLE{Bhattad2007,
  author={Bhattad, Kapil and Narayanan, Krishna R.},
  journal={IEEE Transactions on Information Theory}, 
  title={An MSE-Based Transfer Chart for Analyzing Iterative Decoding Schemes Using a Gaussian Approximation}, 
  year={2007},
  volume={53},
  number={1},
  pages={22-38},
  doi={10.1109/TIT.2006.887074}}

@ARTICLE{Dudeja2024,
  author={Dudeja, Rishabh and Sen, Subhabrata and Lu, Yue M.},
  journal={IEEE Transactions on Information Theory}, 
  title={Spectral Universality in Regularized Linear Regression With Nearly Deterministic Sensing Matrices}, 
  year={2024},
  volume={70},
  number={11},
  pages={7923-7951},
  doi={10.1109/TIT.2024.3458953}}

@ARTICLE{Greig2018,
  author={Greig, Adam and Venkataramanan, Ramji},
  journal={IEEE Transactions on Communications}, 
  title={Techniques for Improving the Finite Length Performance of Sparse Superposition Codes}, 
  year={2018},
  volume={66},
  number={3},
  pages={905-917},
  doi={10.1109/TCOMM.2017.2776937}}

@article{Donoho2009,
  title={Message-passing algorithms for compressed sensing},
  author={David L. Donoho and Arian Maleki and Andrea Montanari},
  journal={Proceedings of the National Academy of Sciences},
  year={2009},
  volume={106},
  pages={18914 - 18919},
}

@INPROCEEDINGS{Donoho2010,
  author={Donoho, David L. and Maleki, Arian and Montanari, Andrea},
  booktitle={2010 IEEE Information Theory Workshop on Information Theory (ITW 2010, Cairo)}, 
  title={Message passing algorithms for compressed sensing: I. motivation and construction}, 
  year={2010},
  volume={},
  number={},
  pages={1-5},
  doi={10.1109/ITWKSPS.2010.5503193}}

@article{Rangan2019vector,
  title={Vector approximate message passing},
  author={Rangan, Sundeep and Schniter, Philip and Fletcher, Alyson K},
  journal={IEEE Transactions on Information Theory},
  volume={65},
  number={10},
  pages={6664--6684},
  year={2019},
  publisher={IEEE}
}

@article{Liu2022memory,
  title={Memory {AMP}},
  author={Liu, Lei and Huang, Shunqi and Kurkoski, Brian M},
  journal={IEEE Transactions on Information Theory},
  volume={68},
  number={12},
  pages={8015--8039},
  year={2022},
  publisher={IEEE}
}

@ARTICLE{Lei2025random,
  author={Liu, Lei and Chi, Yuhao and Huang, Shunqi and Zhang, Zhaoyang},
  journal={IEEE Transactions on Information Theory}, 
  title={Random multiplexing}, 
  year={2026},
  volume={72},
  number={4},
  pages={2277-2306},
  }

@ARTICLE{Barbier2017,
  author={Barbier, Jean and Krzakala, Florent},
  journal={IEEE Transactions on Information Theory}, 
  title={Approximate Message-Passing Decoder and Capacity Achieving Sparse Superposition Codes}, 
  year={2017},
  volume={63},
  number={8},
  pages={4894-4927},
  doi={10.1109/TIT.2017.2713833}}

@INPROCEEDINGS{Liu2025CapacityachievingSS,
  author={Liu, Yuhao and Fu, Teng and Fan, Jie and Niu, Panpan and Deng, Chaowen and Huang, Zhongyi},
  booktitle={2025 IEEE International Symposium on Information Theory (ISIT)}, 
  title={Capacity-Achieving Sparse Superposition Codes with Spatially Coupled VAMP Decoder}, 
  year={2025},
  volume={},
  number={},
  pages={1-6},
}

@ARTICLE{Wang2020,
  author={Wang, Xiaojie and Liang, Chulong and Ping, Li and ten Brink, Stephan},
  journal={IEEE Transactions on Wireless Communications}, 
  title={Achievable Rate Region for Iterative Multi-User Detection via Low-Cost Gaussian Approximation}, 
  year={2020},
  volume={19},
  number={5},
  pages={3289-3303},
  doi={10.1109/TWC.2020.2971999}}

@ARTICLE{Liu2019,
  author={Liu, Lei and Chi, Yuhao and Yuen, Chau and Guan, Yong Liang and Li, Ying},
  journal={IEEE Transactions on Signal Processing}, 
  title={Capacity-Achieving {MIMO-NOMA}: Iterative {LMMSE} Detection}, 
  year={2019},
  volume={67},
  number={7},
  pages={1758-1773},
  doi={10.1109/TSP.2019.2896242}}

@ARTICLE{Chi2022,
  author={Chi, Yuhao and Liu, Lei and Song, Guanghui and Li, Ying and Guan, Yong Liang and Yuen, Chau},
  journal={IEEE Transactions on Communications}, 
  title={Constrained Capacity Optimal Generalized Multi-User {MIMO}: A Theoretical and Practical Framework}, 
  year={2022},
  volume={70},
  number={12},
  pages={8086-8104},
  doi={10.1109/TCOMM.2022.3207813}}

@article{Burak2026,
      title={An Orthogonal Approximate Message Passing Framework for Multiuser Communications}, 
      author={Burak \c{C}akmak and Hao Yan and Alexander Fengler and Giuseppe Caire and Lei Liu},
      year={2026},
      journal={arXiv preprint arXiv:2606.26777},
}

@unpublished{Hao2026,
  author = {Hao Yan and Lei Liu and Yuhao Liu and Burak {\c{C}}akmak and Giuseppe Caire},
  title  = {Capacity-Region-Achieving {MIMO}-{MAC}: Coding Principle and  Sparse Regression Code Design},
  note   = {manuscript in preparation},
  year   = {2026}
}

@article{Ramji2019,
  title={Sparse Regression Codes},
  author={Ramji Venkataramanan and Sekhar Chandra Tatikonda and Andrew R. Barron},
  journal={Found. Trends Commun. Inf. Theory},
  year={2019},
  volume={15},
  pages={1-195},
}

@book{Cover2006,
author = {Cover, Thomas M. and Thomas, Joy A.},
title = {Elements of Information Theory},
year = {2006},
isbn = {0471241954},
publisher = {Wiley-Interscience},
address = {USA},
}

@book{Tse2005,
  title     = {Fundamentals of Wireless Communication},
  author    = {Tse, D. and Viswanath, P.},
  year      = {2005},
  publisher = {Cambridge University Press},
}

@ARTICLE{Fengler2021,
  author={Fengler, Alexander and Jung, Peter and Caire, Giuseppe},
  journal={IEEE Transactions on Information Theory}, 
  title="{SPARCs} for Unsourced Random Access", 
  year={2021},
  volume={67},
  number={10},
  pages={6894-6915},
  }

\end{document}